\documentclass[aps,pra,floatfix,superscriptaddress,twocolumn]{revtex4-1}
\usepackage{amsmath,amssymb}
\usepackage{graphicx}
\usepackage{epstopdf}
\usepackage{bm}
\usepackage{cleveref}
\usepackage{subcaption}
\usepackage{xcolor}

\newcommand{\vk}{\mathbf{k}}

\newcommand{\vbr}{\mathbf{r}}
\newcommand{\be}{\begin{eqnarray}}
\newcommand{\ee}{\end{eqnarray}}

\newcommand{\hx}{\hat{x}}
\newcommand{\hy}{\hat{y}}
\newcommand{\dc}{c^{\dagger}}

\def\ket#1{|#1\rangle}

\def\ep#1{\langle #1 \rangle}

\begin{document}

\title{The Quadrupole Moment of Higher-Order Topological Insulator at Finite temperature}
\author{Yiting Deng}
\affiliation{College of Physics, Sichuan University, Chengdu, Sichuan 610064, China}
\email{heyan$_$ctp@scu.edu.cn}

\author{Yan He}
\affiliation{College of Physics, Sichuan University, Chengdu, Sichuan 610064, China}
\email{heyan$_$ctp@scu.edu.cn}

\begin{abstract}
We study the higher-order topological insulators at finite temperature based on a generalized real-space quadrupole moment, which extends the ground state expectations to ensemble averages. Our study reveals that chiral symmetry alone dictates that the quadrupole moment must be quantized to two values of $0$ and $\pi$, even at finite temperature. It is found that finite temperature can induce a topological phase transition from non-trivial to trivial.
Furthermore, we found that the anisotropic intra-cell hopping can lead to a reentrant topological phase transition, in which the system becomes topological again with rising temperature. This reentrant behavior is in stark contrast to the results at zero temperature.
We also investigate the effects of the quasi-disorder hopping on the topology. It is found that the initially trivial system can be driven into a topological phase with strong enough disorder strength, which closely resembles the topological Anderson transition.
Our work provides an example for studying the finite temperature topology of higher-order topological insulators.
\end{abstract}

\maketitle

\section{Introduction}

In the past decade, the topological phases of matter have become a central topic in both theoretical and experimental condensed matter physics \cite{Kane_TIRev,qi2011topological}. Initially inspired by the quantum Hall effect \cite{QHE,TKNN,haldane}, the studies of topological properties of matter have been greatly enlarged and deepened, and also affect almost all branches of condensed matter physics \cite{Chiu2016,Bernevig_book,shen2012topological}.

A new type of topological phase of matter called the higher-order topological insulators (HOTI) \cite{benalcazar2017quantized,okugawa2019second,benalcazar2022chiral,schindler2018higher,Song-2017,Franca} was proposed a few yeas ago and attracted a lot of attention. The defining feature of the conventional topological insulator is that it can support metallic edge modes localized on the boundary of the system \cite{hatsugai1993chern}. In other words, the edge modes have one dimension less than the bulk states. In the HOTI, the system can support edge modes with two dimensions less than the bulk state.  If we put a 2-dimensional HOTI system on a square with open boundaries along both directions, one can find that there are 4 zero modes located at the corner of the system. For a review of HOTI, one can consult \cite{Yang_2024}.

The most prominent example of the HOTI is the Benalcazar-Bernevig-Hughes (BBH) model, which can be roughly thought of as a two-dimensional tight-binding model with alternating hopping constants \cite{benalcazar2017quantized}. The conventional topological insulator can be described by the Berry phase, which is also equivalent to the electric dipole moment.
Inspired by this, it was suggested that the topological property of the HOTI on a square can be characterized by the quadrupole moment $q_{xy}$. Therefore, it is also known as the topological quadupole insulator (TQI). In the original work of the BBH model, the authors proposed that one can compute the topological index of a HOTI by the method of nested Berry phase. Later, it is found that the quadrupole moment $q_{xy}$ can also be directly computed in real space \cite{kang2019many,wheeler2019many}. Although these two approaches agree with each other, we will mainly follow the real space calculations of $q_{xy}$ in the rest of this paper.

Ever since the pioneer work of the BBH model, many researchers have extended the notion of HOTI to include a variety of different effects, such as disorders \cite{li2020topological,yang2021higher}, quasi-disorders \cite{zeng2020higher,lahiri2024quasiperiodic}, or quasicristals \cite{peng2021higher,li2022transition}. Most of these studies primarily focused on the TQI at zero temperature. However, in the real physical world, the finite temperature effects or the thermal fluctuations will inevitably affect the properties of the system. The primary goal of this paper is to understand how to characterize the topology of TQI at finite temperatures.

There have already been many works that extend the topological indices at $T=0$ to finite $T$, such as the Uhlmann phase \cite{Viyuela14,Huang14} or the ensemble geometric phases (EGP)\cite{Bardyn_2018}.
The Uhlmann phase is based on the Uhlmann connection, which is the counterpart of the Berry connection for mixed states \cite{Viyuela15}.
The EGP, on the other hand, represents a direct extension of the ground-state expectation values of the electric polarization to ensemble averages. So far, these methods have mainly been applied to study conventional topological insulators. In this work, we will closely follow the idea of EGP and generalize the quadrupole moments at $T=0$ to ensemble averages. We will show that this finite $T$ version of the quadrupole moment is still quantized if the system has chiral symmetry. Due to this quantization, we propose to use this generalized quadrupole moment as a topological index to investigate the TQI at finite $T$. Equipped with this new tool, we can find out how the temperature affects the topological phase transition in TQI. We would like to mention that the finite temperature topology of HOTI is also considered in \cite{lu2025finite}.

The rest of this paper is organized as follows. In \cref{sec:T0}, we briefly review the BBH model and the quadrupole moment $q_{xy}$ at zero temperature. In \cref{sec:FT}, we generalize the quadrupole moment to finite temperature and demonstrate that the quadrupole moment is quantized for systems with chiral symmetry. The numerical results of the quadrupole moment at finite temperature for the BBH model with both isotropic and anisotropic hopping are shown in \cref{sec:num}. In this section, we also study how the quasi-periodic hopping affects the topological phase transitions of TQI. Finally, a summary is presented in \cref{sec:conclu}.

\section{The quadrupole moment at zero temperature}
\label{sec:T0}

In this section, we briefly review the topological properties of the BBH model \cite{benalcazar2017quantized} with chiral symmetry at zero temperature. We also discuss how to compute the topological invariants that characterize the topology of this model.

The Hamiltonian of the BBH model in real space is given as
\be
&&\mathcal{H}=\sum_{\vbr}\Big[t_x(\dc_{\vbr,1}c_{\vbr,3}+\dc_{\vbr,4}c_{\vbr,2})+t_y(\dc_{\vbr,1}c_{\vbr,4}-\dc_{\vbr,3}c_{\vbr,2})\nonumber\\
&&\qquad +t(\dc_{\vbr,1}c_{\vbr+\hx,3}+\dc_{\vbr,4}c_{\vbr+\hx,2})\nonumber\\
&&\qquad +t(\dc_{\vbr,1}c_{\vbr+\hy,4}-\dc_{\vbr,3}c_{\vbr+\hy,2})+H.c.\Big]
\label{eq-Ham}
\ee
Here $c_{\vbr,a}^{\dagger}$ and $c_{\vbr,a}$ are the fermion creation and annihilation operators located at the lattice site $\vbr=(x,y)$ with $x=1,\cdots,L_x$ and $y=1,\cdots,L_y$. The orbital index $a=1,2,3,4$ denotes the four orbitals in each unit cell. We also use $\hx$ and $\hy$ to represent the unit vectors along the $x$ and $y$ directions.

If we impose a periodic boundary condition, the BBH model can be simplified in the momentum space, which reads as follows
\be
&&\mathcal{H}=\sum_{\vk}\psi^{\dag}_\vk H(\vk)\psi_\vk\nonumber\\
&& H(\vk)=(t_x+t\cos k_x)\Gamma_4+t\sin k_x\Gamma_3\nonumber\\
&&\quad +(t_y+t\cos k_y)\Gamma_2+t\sin k_y\Gamma_1
\ee
where $\psi_\vk=(c_{\vk,1},\,c_{\vk,2},\,c_{\vk,3},\,c_{\vk,4})^T$. The $\Gamma$ matrices are defined as $\Gamma_{j}=-\tau_{2}\sigma_{j}$ for $j=1,2,3$ and $\Gamma_{4}=\tau_{1}\sigma_{0}$. Here $\tau$ and $\sigma$ are both the Pauli matrices acting on different orbital degrees of freedom. The BBH model has chiral symmetry, which means that the Hamiltonian satisfies the anti-commuting relation $\{\Pi,\mathcal{H}(\vk)\}=0$. The chiral symmetry operator is given as $\Pi=\tau_{3}\sigma_{0}$.

To describe the topology, we introduce a real-space formula for the electric quadrupole moment at $T=0$.  It is proven that this quadrupole moment is quantized when the system has chiral symmetry. Due to this property, one can treat the quadrupole moment as a topological index for a TQI, such as the BBH model. The real-space quadrupole moment was proposed in the works \cite{kang2019many,wheeler2019many} and was further developed in \cite{li2020topological,yang2021}. It can be thought of as a generalization of the Resta formula \cite{resta1998quantum} for electric polarization.

The quadrupole moment operator is usually defined as
\be
\hat{Q}_{xy}=\sum_{j=1}^N\frac{x_j y_j}{L_x L_y}\dc_j c_j
\ee
Here $j=(\vbr,a)$ is a collective index denoting both the lattice site and the orbital in each unit cell. $x_j$ and $y_j$ are the coordinate of the $j$th orbital. The total number of orbital in the whole system is $N=4L_xL_y$. However the above definition of the quadrupole moment is not compatible with the periodic boundary condition. Inspired by the Resta formula of charge polarization, one can define the quadrupole moment as
\be
q_{xy}=\frac1{2\pi}\arg\ep{G|\exp(2\pi i\hat{Q}_{xy})|G}-q^0_{xy}
\ee
Here arg$(\cdots)$ represents taking the argument or phase angle of the following quantity. $\ket{G}$ is the many-body ground state of the BBH model. In the above definition, we have subtract a background charge quadrupole $q^0_{xy}$ which is obtained by assuming a half-filled fermion number for all unit cell. More explicitly, the background quadrupole can be computed as
\be
q^0_{xy}=\frac12\sum_{j=1}^N\frac{x_jy_j}{L_xL_y}\mod 1
\ee
It is easy to see that the above defined $q_{xy}$ satisfies $0\le q_{xy}\le1$.

In the first quantized language, the ground state $\ket{G}$ can be represented by a $N\times N_{occ}$ matrix $U_o$ whose columns are the occupied eigenstates of $H$. More explicitly, the matrix $U_o$ reads as follows
\be
U_{o}=\Big\{\ket{\psi_1},\,\ket{\psi_2},\cdots,\ket{\psi_{N_{occ}}}\Big\}
\ee
with $N=4L_xL_y$ as the total number of states and $N_{occ}$ as the total number of occupied states.

In the first quantized language, the operator $e^{2\pi i\hat{Q}_{xy}}$ can be expressed as a diagonal matrix as follows
\be
D=\mbox{diag}\Big\{\exp\big(2\pi i\frac{x_j y_j}{L_x L_y}\big)\Big\}
\label{eq-D}
\ee
with $j=1,\cdots,N$.
Then the above expectation value can be computed by a determinant as follows
\be
\ep{G|\exp\Big(2\pi i\frac{x y}{L_x L_y}\hat{n}_\vbr\Big)|G}=\det(U_o^{\dagger}D U_o)
\ee
Note that the background quadrupole can also be rewritten as $q^0_{xy}=\frac1{2\pi}\arg[\det(D^{1/2})]$ where $\det(D^{1/2})=\prod_{j=1}^N\exp\big(\pi i\frac{x_jy_j}{L_x L_y}\big)$. Taking into account of this term, one find that $q_{xy}$ can be expressed as
\be
q_{xy}=\frac{1}{2\pi}\arg\Big[\det(U_o^{\dagger}D U_o)\frac1{\det(D^{1/2})}\Big]
\label{qxy-T0}
\ee
which agrees with the result in \cite{li2020topological,Yang_2024}.
We mention that the calculation of the above formula is under the periodic boundary condition.

\begin{figure}[h]
\begin{center}
\includegraphics[width=0.9\columnwidth]{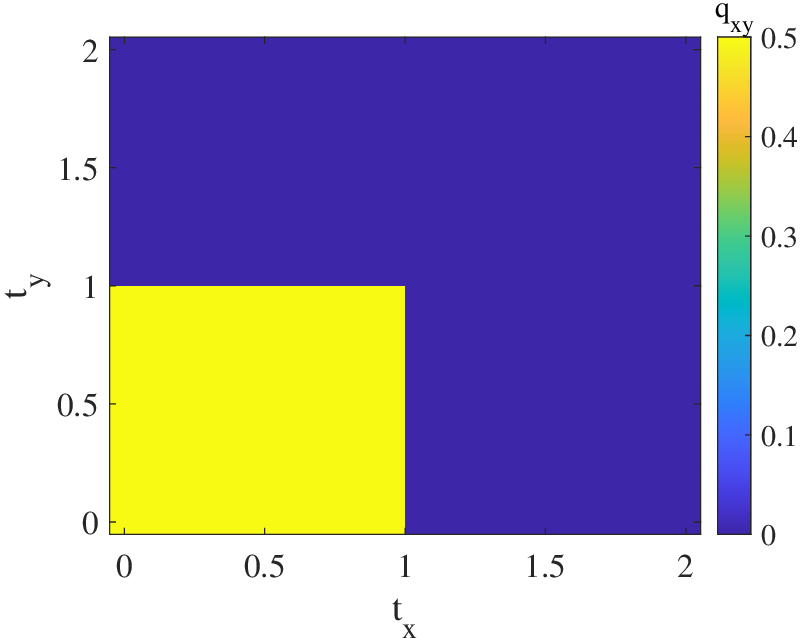}
\caption{The quadrupole moment $q_{xy}$ of the BBH model as a function of the $t_{x}$ and $t_{y}$. Here we also assume that $t=1$ and the system size is $L=30$.}
\label{fig-pd-T0}
\end{center}
\end{figure}

As an example of the application of Eq.~(\ref{qxy-T0}), we plot the quadrupole moment $q_{xy}$ as a function of the $t_{x}$ and $t_{y}$ in the \cref{fig-pd-T0}, One can see that there are two distinct phases that can be clearly identified: one with $q_{xy} = 0.5$ (depicted in yellow) when $t_{x} < 1$ and $t_{y} < 1$, indicating a nontrivial higher-order topological phase; and the other with $q_{xy} = 0$ (shown in blue) when $t_{x} > 1$ or $t_{y}> 1$, representing a trivial normal insulator phase.

\section{The quadrupole moment at finite temperature}
\label{sec:FT}

In this section, we extend the definition of the quadrupole moment from zero temperature to finite temperature. We will see that this generalized quadrupole moment will allow one to investigate the topology of HOTI at finite temperature.

At finite temperature, a straightforward generalization is to replace the ground state expectation values by an ensemble average. Therefore, the finite $T$ quadrupole moment is defined as
\begin{equation}
q_{xy}=\frac{1}{2\pi}\arg\Big[\mbox{Tr}(\rho e^{2\pi i\hat{Q}_{xy}})\Big]-q_{xy}^0
\label{eq-defin}
\end{equation}
The density matrix of the system is given by
\be
\rho=\frac{1}{Z}\exp\Big(-\beta\sum_{i,j=1}^N c_{i}^{\dagger}H_{ij}c_{j}\Big)
\ee
where $H$ is the first-quantized Hamiltonian in the coordinate basis. Here $\beta=\frac{1}{k_{B}T}$ and for convenience we assume $k_{B}=1$.

The partition function $Z$ is computed as
\begin{equation}
Z=\mbox{Tr}\Big[\exp(-\beta\sum_{i,j}c_{i}^{\dagger}H_{ij}c_{j})\Big]=\det(I+e^{-\beta H})
\label{eq-Z}
\end{equation}
with $I$ being the $N$ by $N$ identity matrix.

Making use of the following identity
\be
\mbox{Tr}\Big(e^{\sum_{i,j}\dc_iA_{ij}c_j}e^{\sum_{i,j}\dc_iB_{ij}c_j}\Big)
=\det(I+e^Ae^B)
\ee
which is proved in \cite{Bardyn_2018,pi2022proxy}, we can compute the trace in Eq.~(\ref{eq-defin}) as
\begin{equation}
\mbox{Tr}\Big(\rho e^{2\pi i\hat{Q}_{xy}}\Big)=\frac{1}{Z}\det(I+e^{-\beta H}D)
\label{eq-Tr}
\end{equation}
Substituting Eq.~(\ref{eq-Z}) and (\ref{eq-Tr}) into Eq.~(\ref{eq-defin}), we finally arrived at the following result
\begin{equation}
q_{xy}=\frac{1}{2\pi}\arg\Big[\frac{\det(I+e^{-\beta H}D)}{\det(I+e^{-\beta H})\det(D^{1/2})}\Big]
\label{qxy-ft}
\end{equation}
This formula is a central result of this paper.

Next, we will demonstrate that the finite $T$ quadrupole moment of Eq.~(\ref{qxy-ft}) will remains quantized for systems with chiral symmetry. More explicitly, the value of $q_{xy}$ continues to take the value of $0$ or $1/2$. Subsequently, we will demonstrate that at zero temperature, Eq.~(\ref{qxy-ft}) reduce to the corresponding zero temperature quadrupole moment as the consistency requires. Furthermore, at the infinite high temperature, the value of $q_{xy}$ is bound to be zero, as one usually expected.

\subsection{Quantization of the $q_{xy}$ at finite $T$ with Chiral Symmetry}

First, we show that $q_{xy}$ can only take two quantized values $0$ or $1/2$ for a system with chiral symmetry.
According to Eq.~(\ref{qxy-ft}), $q_{xy}$ is quantized if and only if the three determinant inside the bracket are real numbers.
Let us first look at the denominators. Assuming $E_j$ with $j=1,\cdots, N$ are the eigenvalues of the Hamiltonian $H$, then we have
\be
\det(I+e^{-\beta H})=\prod_{j}(1+e^{-\beta E_{j}})
\ee
Since all eigen-energy $E_j$ are real numbers, it is easy to see that the above determinant is a positive number.
Recall that there are 4 orbital in each unit cell, then the determinant $\det(D^{1/2})$ can computed as
\be
& &\det(D^{1/2})=\exp\Big(\pi i\sum_j\frac{x_j y_j}{L_x L_y}\Big)\nonumber\\
&=&\exp\Big(4\pi i\sum_{n_x=1}^{L_x}\sum_{n_y=1}^{L_y}\frac{n_x n_y}{L_x L_y}\Big)\nonumber\\
&=&(-1)^{(L_x+1)(L_y+1)}
\ee
which is clearly equals to $\pm1$. Now we turn to the numerator. Note that according to the definition of the matrix $D$ in Eq.~(\ref{eq-D}), we have $D^*=D^{-1}$, where the symbol $^*$ means taking a complex conjugate. Recall that the chiral symmetry requires that $\{\Pi,H\}=0$ with $\Pi^2=1$. Making use of these facts, we find that
\be
& &\det(I+e^{-\beta H}D)^{*}=\det(I+e^{-\beta H}D^{-1})\nonumber\\
&=&\det\Big[\Pi(I+e^{-\beta H}D^{-1})\Pi\Big]=\det(I+e^{\beta H}D^{-1})\nonumber\\
&=&\det(e^{\beta H}D^{-1})\det(I+e^{-\beta H}D)
\label{eq-det}
\ee
In the second line of above equation, we have used the fact that $[\Pi,D]=0$. It is easy to see that
\be
\det D=\det D^{-1}=e^{i2\pi(L_x+1)(L_y+1)}=1.
\ee
Making use of chiral symmetry, we also have
\be
\det(e^{-\beta H})=\det(\Pi e^{-\beta H}\Pi)=\det(e^{\beta H})
\ee
which implies that $\det(e^{-\beta H})=\det(e^{\beta H})=1$. Applying the above results in Eq.~(\ref{eq-det}), we find that
\be
\det(I+e^{-\beta H}D)^{*}=\det(I+e^{-\beta H}D)
\ee
which implies that $\det(I+e^{-\beta H}D)$ is a real number.

Based on the above analysis, we find that the quantity inside the bracket of Eq.~(\ref{qxy-ft}) is a real number. Therefore, one can conclude that the value of $q_{xy}$ must be restricted to either $0$ or $1/2$.

In this paper, we will always assume that $L_x=L_y=L$ where $L$ is an even integer. In this case, we always have $\det(D^{1/2})=-1$. Note that $\det(I+e^{-\beta H})$ is always positive, then by Eq.~(\ref{qxy-ft}), we find a simpler formula for the quadupole moment as follows
\be
q_{xy}=\left\{
         \begin{array}{ll}
           1/2, &  \det(I+e^{-\beta H}D)>0\\
           0, & \det(I+e^{-\beta H}D)<0
         \end{array}
       \right.
       \label{eq-qxy2}
\ee

\subsection{The zero Temperature limit of $q_{xy}$}

Now we consider the zero temperature limit of $q_{xy}$ and show that it can go back to the result of Eq.(\ref{qxy-T0}).
It is well-known that the Hamiltonian can be diagonalized as $U^{\dag}HU=\Lambda$. Here $U$ is an $N\times N$ Unitary matrix whose columns are eigenstates of $H$. It can be decomposed as $U=\{U_o,\,U_u\}$, where $U_o$ is made by the first $N_{occ}$ columns of occupied eigenstates and $U_u$ is made by the rest $N-N_{occ}$ columns of unoccupied eigenstates.

Correspondingly, the diagonal matrix $\Lambda$ can also be decomposed as $\lambda=\mbox{diag}\{\Lambda_o,\,\Lambda_u\}$. We have defined $\Lambda_o=\mbox{diag}\{E_1,\cdots,E_{N_{occ}}\}$ with $E_j<0$ for $j=1,\cdots,N_{occ}$. These eigenvalues correspond to the occupied eigenstates. On the other hand, $\Lambda_u=\mbox{diag}\{E_{N_{occ}+1},\cdots,E_{N}\}$ corresponds to the un-occupied eigenstates. Note that as $\beta\to\infty$, we find that $e^{-\beta E_j}$ become very large if $E_j<0$ and $e^{-\beta E_j}\approx0$ if $E_j>0$.

With the above preparations, we are ready to observe that the denominator of Eq.~(\ref{qxy-ft}) becomes the following as $T\to0$
\begin{equation}\begin{aligned}
&\det(I+e^{-\beta H})=\det(I+e^{-\beta\Lambda})\\
&=\det(I+e^{-\beta \Lambda_{o}})\approx\det e^{-\beta \Lambda_{o}}
\label{mixed state fenmu}
\end{aligned}\end{equation}
By diagonalizing $H$, we also find that
\be
& &\det(I+e^{-\beta H}D)=\det(I+e^{-\beta\Lambda}U^{\dagger}DU)\nonumber\\
&=&\det\Bigg[\left(\begin{array}{cc}
I_{o} & 0\\
0 & I_{u}
\end{array}\right)+\left(\begin{array}{cc}
e^{-\beta \Lambda_o} & 0\\
0 & 0
\end{array}\right)\left(\begin{array}{cc}
U_{o}^{\dagger}DU_{o} & U_{o}^{\dagger}DU_{u}\\
U_{u}^{\dagger}DU_{o} & U_{u}^{\dagger}DU_{u}
\end{array}\right)\Bigg]\nonumber\\
&=&\det(I_{o}+e^{-\beta \Lambda_{o}}U_{o}^{\dagger}DU_{o})\nonumber\\
&\approx&\det(e^{-\beta \Lambda_{o}}U_{o}^{\dagger}DU_{o})
\label{mixed state:fenmz}
\ee
where $I_o$ and $I_u$ are identity matrices with dimension $N_{occ}$ and $N-N_{occ}$ respectively.
Combining the results of the above two equations, we find that
\begin{equation}\begin{aligned}
\frac{\det(I+e^{-\beta H}D)}{\det(I+e^{-\beta H})}=\det(U_{o}^{\dagger}DU_{o})
\label{det-UDU}
\end{aligned}\end{equation}
From the above equation, it clear that Eq.~(\ref{qxy-ft}) reduced to Eq.~(\ref{qxy-T0}) as $T\to0$. Therefore, the definition of quadrupole moment $q_{xy}$ at finite $T$ is consistent the $q_{xy}$ defined at $T=0$.

\subsection{The infinite Temperature limit of $q_{xy}$}

As the temperature $T$ approaches to infinity, the matrix $e^{-\beta H}$ approaches to an identity matrix. Therefore, we find that
\be
&&\lim_{\beta\to0}\det(I+e^{-\beta H}D)=\det(I+D)\nonumber\\
&&\quad =\prod_{j=1}^N\Big(1+e^{2\pi i\frac{x_jy_j}{L_{x}L_{y}}}\Big)
\ee
Assuming $L_x=L_y=L$, then both $x_j$ and $y_j$ takes values among $\{{1,\cdots,L}\}$. Recall that each unit cell contains 4 orbital, we thus find that
\be
&&\prod_{j=1}^N\Big(1+e^{i2\pi\frac{x_jy_j}{L_{x}L_{y}}}\Big)
=\prod_{n_x,n_y=1}^L\Big[1+\exp(2\pi i\frac{n_x n_y}{L^2})\Big]^4\nonumber\\
&&\quad=\prod_{n_x,n_y=1}^L2^4\cos^4(\pi\frac{n_x n_y}{L^2})\exp(4\pi i\frac{n_x n_y}{L^2})
\ee
If we assume that $L$ is an even integer, then we find that
\be
&&\prod_{n_x,n_y=1}^N\exp(4\pi i\frac{n_x n_y}{L^2})=\exp(4\pi i\frac{L^2(L+1)^2}{4L^{2}})\nonumber\\
&&\quad =e^{i\pi(L+1)^{2}}=-1
\ee
Note that $\cos^4(\pi\frac{n_x n_y}{L^2})$ is positive, Therefore we conclude that
\be
\det(I+D)<0
\ee
which means $q_{xy}=0$ according to Eq.~(\ref{eq-qxy2}). Therefore, in the limit of infinite high temperature, the system becomes topologically trivial.

\section{Numerical Results and Discussion}
\label{sec:num}

\begin{figure}[h]
\begin{center}
\includegraphics[width=0.9\columnwidth]{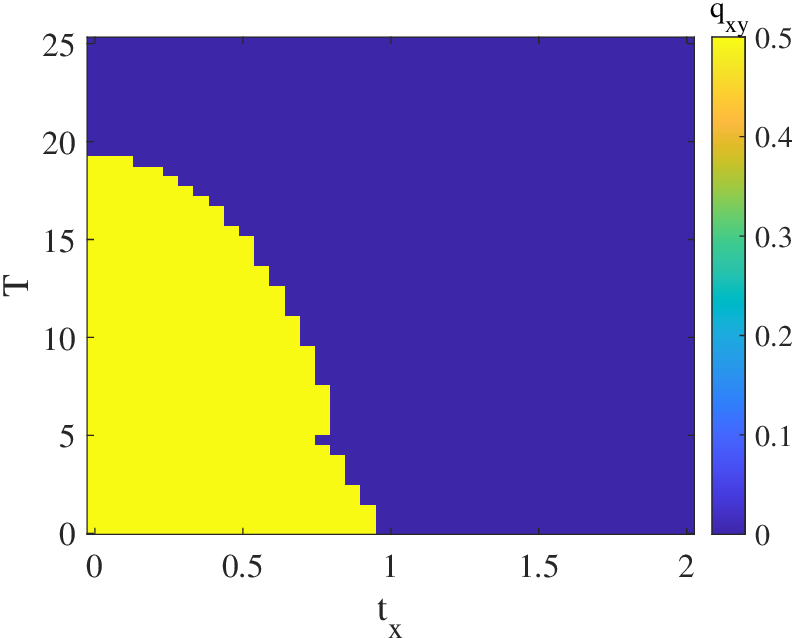}
\caption{The quadrupole moment $q_{xy}$ of isotropic BBH model as the function of $t_{x}$ and $T$. Here we assume that $t_x=t_y$, $t=1$ and $L=30$.}
\label{fig-pd-iso}
\end{center}
\end{figure}

In this section, we will treat the finite temperature quadrupole moment $q_{xy}$ of Eq.~(\ref{qxy-ft}) as a topological index and apply it to investigate the topological properties of the BBH model or TQI. We first consider the BBH model with isotropic intra-cell hoppings, which means we always assume $t_x=t_y$ in this case. The quadrupole moment $q_{xy}$ as a function of $t_{x}$ and $T$ is shown in the Fig. \ref{fig-pd-iso}. From this phase diagram, we can see that there are still two regions, where the yellow region corresponds to the topological phase with $q_{xy}=1/2$, and the blue region corresponds to the trivial phase with $q_{xy}=0$. One can see that $q_{xy}$ only takes two possible values $0$ and $1/2$, which is consistent with the discussion in the Fig. \ref{sec:FT} that $q_{xy}$ is quantized.

In addition, Fig. \ref{fig-pd-iso} shows that the critical temperature $T_{c}$ between the topologically nontrivial phase and the trivial phase decrease as the hopping constant $t_{x}$ increases. As the $T_c$ approaches zero, the phase boundary approaches $t_x=t_y=1$, which is consistent with the phase diagram at $T=0$. This suggests that the non-zero temperature or thermal fluctuations always suppress the nontrivial topology. For a given $t_{x}<1$, we know the system is topological at $T=0$. As we discussed in Fig. \ref{sec:FT}, the system is definitely topologically trivial at the infinite high $T$. Therefore, there inevitably exists a critical $T_c$ separating the topological and trivial phases for any $0<t_x<1$.

\begin{figure}[!htbp]
\centering
\includegraphics[width=0.9\columnwidth]{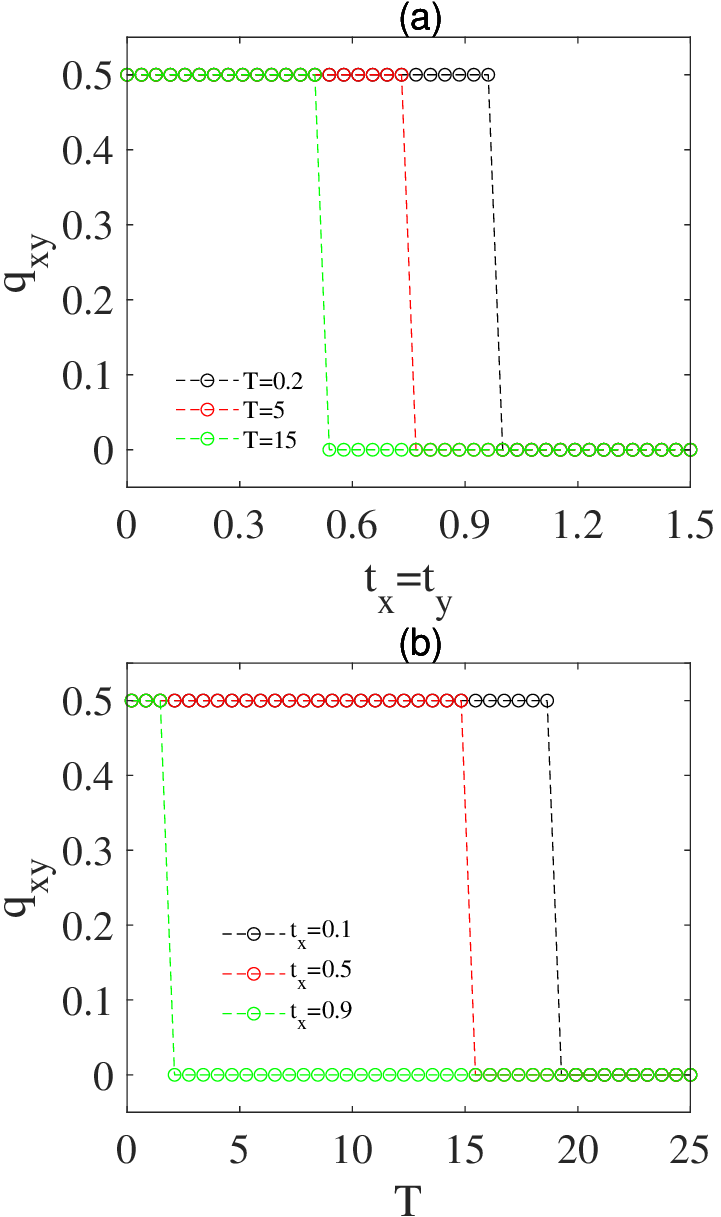}
\caption{(a): The quadrupole moment $q_{xy}$ as a function of $t_x$ for $T=0.3,\,5,\,15$. (b): The quadrupole moment $q_{xy}$ as a function of the $T$ for $t_x=0.1,\,0.5,\,0.9$. In both panels, we assume $t_x=t_y$, $t=1$ and $L=30$.}
\label{fig-iso}
\end{figure}

To get a closer look, we plot $q_{xy}$ as a function of $t_x$ for a few selected temperatures in Fig. \ref{fig-iso} (a). The curves of $T=0.2,\,5,\,15$  are represented by the black, red, and green circles, respectively. At $t_x=0$, all three temperatures satisfy $T<T_c$; thus, all three curves start with $q_{xy}=1/2$. The increase of $t_{x}$ induces an abrupt transition in $q_{xy}$ from $1/2$ to $0$. Furthermore, at lower temperatures, the parameter $q_{xy}$ maintains its nontrivial value of $1/2$ for a larger range of $t_{x}$. For $T=0.2$, the phase boundary is located at $t_{x}\approx1$, which agrees with the phase boundary at $T=0$ of Fig. \ref{fig-pd-T0}. If we consider a case with $T$ larger than the $T_c$ at $t_x=0$, then $q_{xy}$ will remain identically zero for all $t_{x}$ (Not shown in Fig. \ref{fig-iso}).

Similarly, in Fig. \ref{fig-iso} (b), we display the $q_{xy}$ as a function of temperature $T$ for a few selected values of $t_{x}$, specifically $t_{x}=0.1,\,0.5,\,0.9$ corresponding to the colors of black, red, green, respectively. The result shows that for $t_{x}<1$, $q_{xy}$ exhibits a sharp jump from $1/2$ to $0$ as temperature increases, indicating a topological phase transition from a non-trivial to a trivial phase. Moreover, the $T_c$ increases as $t_{x}$ becomes smaller. This suggests that for smaller values of $t_{x}$, the system requires a higher temperature to suppress its nontrivial topology.

In the above calculations, we choose the system size of $L=30$ in both the $x$ and $y$ directions. Although the size of $L=30$ is far different from the thermodynamic limit, we have checked that further increasing the system size will only result in a small change of the phase boundary. Due to this reason, we will stick to the size of $L=30$ for the following numerical computations.

\begin{figure}[h]
\begin{center}
\includegraphics[width=0.9\columnwidth]{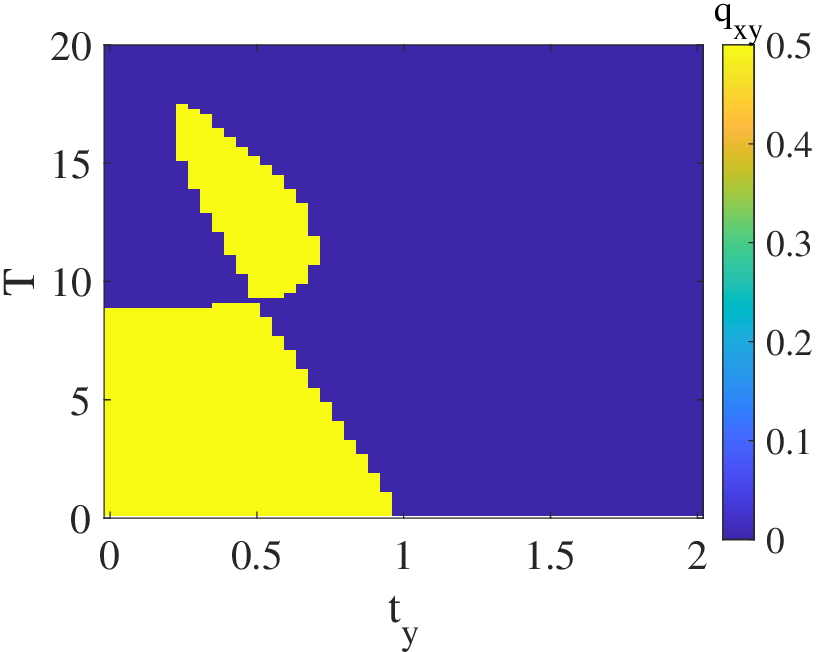}
\caption{The quadrupole moment $q_{xy}$ of isotropic BBH model as the function of $t_y$ and $T$. Here we assume that $t_x=0.5$, $t=1$ and $L=30$.}
\label{fig-pd-aniso}
\end{center}
\end{figure}

\begin{figure}[!htbp]
\centering
\includegraphics[width=0.9\columnwidth]{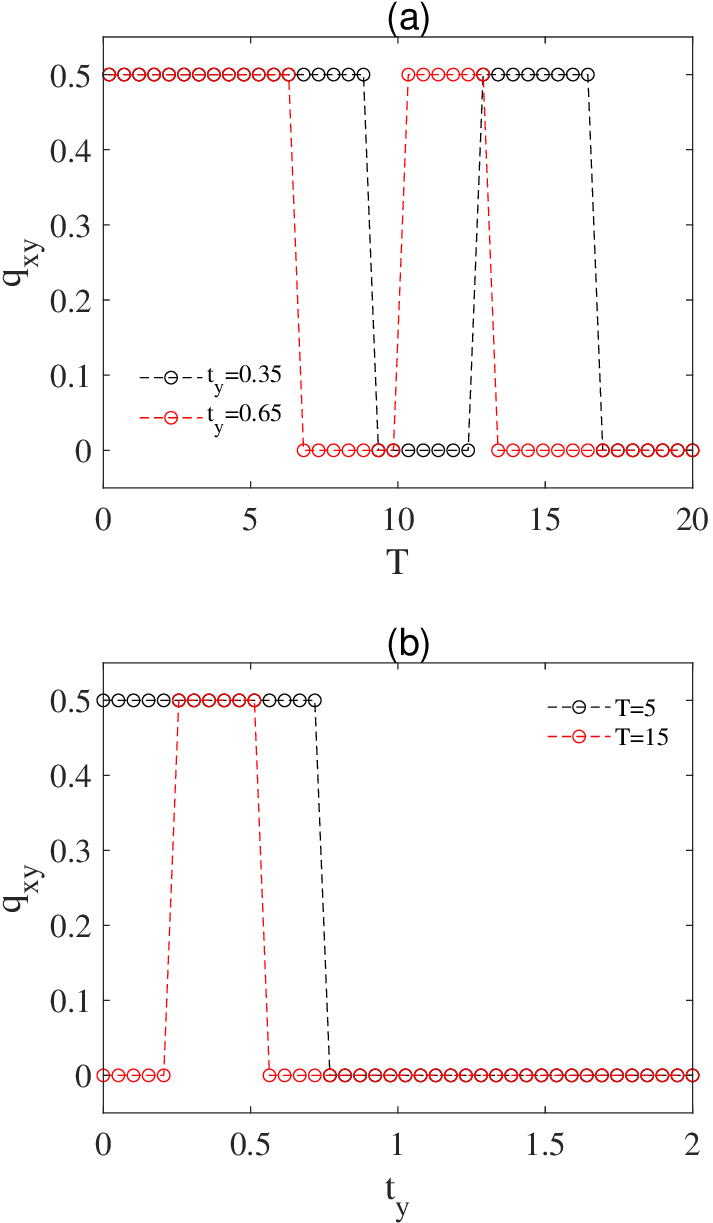}
\caption{(a): The quadrupole moment $q_{xy}$ as a function of $T$ for $t_y=0.35,\,0.65$. (b): The quadrupole moment $q_{xy}$ as a function of the $t_y$ for $T=5,\,15$. In both panels, we assume $t_x=0.5$, $t=1$ and $L=30$.}
\label{fig-aniso}
\end{figure}

Next, we turn to an anisotropic BBH model with different intra-cell hopping $t_{x}\neq t_{y}$.  As a specific example, we set $t_{x}=0.5$ without loss of generality. In Fig. \ref{fig-pd-aniso}, we display the $q_{xy}$ as a function of the temperature $T$ and the intra-cell hopping $t_{y}$. Again, the yellow region represents $q_{xy}=1/2$, which is a topological phase. The blue region with $q_{xy}=0$ is the trivial phase.
At the low $T$ limit, the phase boundary is located at $t_x=1$, which is consistent with Fig. \ref{fig-pd-T0}.  The quadrupole moment of this anisotropic BBH model is still quantized because the chiral symmetry is still intact.
But the phase diagram shows a small island of topological phase in the high $T$ area, which is quite different from the isotropic case. Due to the existence of this island, there will be a reentrant topological phase transition.

To get a better view of this reentrant effect, we take $t_{y}=0.35$ (black circle) and $t_y=0.65$ (red circle) as examples to show $q_{xy}$ as a function of $T$ in Fig. \ref{fig-aniso} (a). For these two fixed $t_y$, the $q_{xy}$ of the system will first jump from $1/2$ to $0$ at a certain intermediate temperature, then jump back to a nontrivial value as temperature continues to rise. Thus, we find a high $T$ topological region, which can sustain for a certain temperature range. But as $T$ keeps increasing, $q_{xy}$ eventually goes back to zero and the system becomes topologically trivial. We can call this transition from trivial to high-$T$ topological phase a reentrant phase transition. The existence of this high-$T$ topological phase is a surprise, since we usually think that the thermal fluctuations will suppress topology. We suspect the unequal intra-cell hopping constants must play an important role in creating a trivial region inside the topological phase. Similarly, we show the $q_{xy}$ as a function of $t_y$ in Fig. \ref{fig-aniso}. For $T=15$, the system is trivial for $t_y\sim0$. While the increase of $t_y$ can drive the system into topological phase.

\begin{figure*}
\centering
\includegraphics[width=\textwidth]{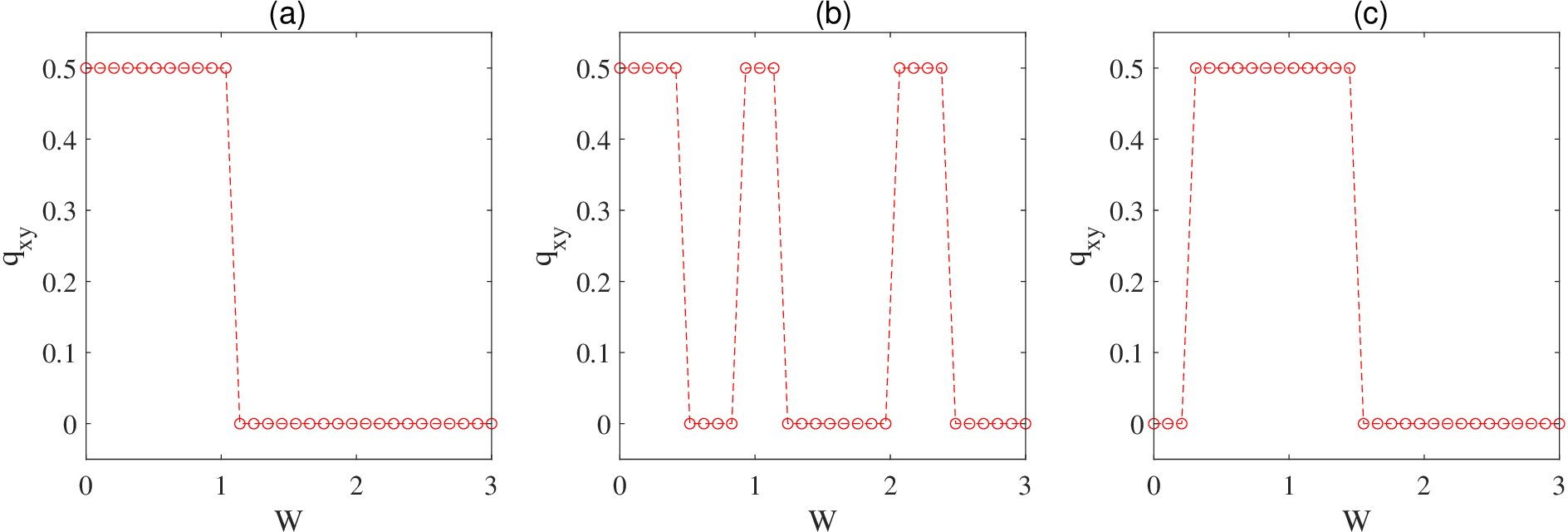}
\caption{The quadrupole moment $q_{xy}$ as a function of quasi-disorder strength $W$ for the quasi-disordered BBH model. From panel (a) to (c), the corresponding temperatures are $T=3$, $3.9$ and $4.5$, respectively. Other parameters are $t=1$ and $L=30$.}
\label{fig-quasi}
\end{figure*}

At last, we want to consider how the quasi-disorder affects the nature of topological phase transitions in the BBH model. To this end, we introduce quasi-disordered intra-cell hopping along the $x$ direction of the BBH model. The quasi-disordered hopping can be expressed as follows:
\begin{equation}\begin{aligned}
t_{n}=t_{x}+W\cos(2\pi\alpha x_{n})
\label{quasidisorder}
\end{aligned}\end{equation}
where $n$ labels the lattice site along the $x$-direction and $\alpha$ is an irrational number which is assumed to be the golden ratio $(\sqrt{5}+1)/2$.
Note that the chiral symmetry of the BBH model is still intact even with this quasi-disordered hopping.

In Fig. \ref{fig-quasi}, we display how the quadrupole moment $q_{xy}$ varies with increasing quasi-periodic disorder strength $W$. In the panel (a) with $T=3$, it is observed that $q_{xy}=1/2$ in the clean limit. As the quasi-disorder strength $W$ increases, $q_{xy}$ persists at the nontrivial value for a while, then drops directly to zero near $W=1.1$. This phenomenon suggests that the quasi-disorder can induce a topological phase transition in TQI, leading the system from a topologically nontrivial phase to a trivial phase. On the other hand, in the panel (c) with $T=4.5$, the system is in a trivial phase at the clean limit. The quasi-disorder induces a rise in the quadrupole moment from $0$ to $1/2$. The system is topologically non-trivial in the region of $0.3<W<1.5$. Then it becomes trivial again for even larger $W$. This suggests that the quasi-disorders can drive the system into a topological phase, which is similar to the topological Anderson insulator. We also plot the $q_{xy}$ curve at an intermediate temperature $T=3.9$ in panel (b). One can see that at this temperature, $q_{xy}$ is jumping between $0$ and $1/2$ several times, exhibiting oscillatory behavior around $0.5<W<2.5$. Only when $W$ becomes even larger does the system go back into the normal insulator phase. Around $T=3.9$, the oscillatory behavior of $q_{xy}$ is typical. We feel that this behavior might reflect certain instability due to the competition between the finite $T$ effect and the quasi-disorders.

\section{Conclusion}
\label{sec:conclu}

In this paper, we have proposed a generalized quadrupole moment for mixed states, with close analogy to the concept of EGP.
We rigorously demonstrate that the quadrupole moment of a TQI at finite $T$ should be quantized if the system possesses chiral symmetry. At an infinite high temperature, it is shown that the quadrupole moment must be zero. Thus, the system will undergo a topological phase transition upon increasing the temperature, if it is in the nontrivial phase at $T=0$.  Our later numerical results confirmed that the thermal fluctuation indeed suppresses the topology as expected. Additionally, we observed in a TQI with anisotropic intra-cell hopping that the system will first become trivial and then turn back to non-trivial again as the temperature keeps increasing. This phenomenon is what we call a reentrant phase transition, which is quite generic in a TQI model with unequal hopping constants. At last, we also show that the quasi-disorder hopping in a HOTI model can drive the system from a trivial phase to a topological phase at finite temperature, which is similar to the Topological Anderson transition.

\begin{acknowledgments}
This work is supported by the National Natural Science Foundation of China under Grant No. 11874272.
\end{acknowledgments}


%

\end{document}